\documentclass[11pt,a4paper]{article}
\usepackage[utf8]{inputenc}
\usepackage{bm}
\usepackage{url}
\usepackage{parskip}
\usepackage{amsmath}
\usepackage{amssymb}
\usepackage{xcolor}
\usepackage{fancyvrb}
\usepackage{framed}
\usepackage{authblk,url}
\usepackage{float}
\usepackage{cite}
\usepackage{bbm}
\usepackage{enumerate}
\usepackage{graphicx}
\usepackage{multirow}

\begin{document} 

\title{Calculating Expected Value of Sample Information Adjusting for Imperfect Implementation}

\author{Anna Heath}

\date{May 2021}

\maketitle

\section*{Abstract}

\textbf{Background}: The Expected Value of Sample Information (EVSI) calculates the value of collecting additional information through a study with a given design. Standard EVSI analyses assume that the treatment recommendations based on the new information will be implemented immediately and completely once the study has finished. However, treatment implementation is often slow and incomplete, giving a biased estimation of the study value. Previous methods have adjusted for this bias, but they typically make the unrealistic assumption that the study outcomes do not impact the implementation. One method does assume that the implementation is related to the strength of evidence in favour of the treatment but this method uses analytical results, which require alternative restrictive assumptions.

\textbf{Methods}: We develop two implementation-adjusted EVSI calculation methods that relax these assumptions. The first method uses computationally demanding nested simulations, based on the definition of the implementation-adjusted EVSI. The second method aims to facilitate the computation by adapting a recently developed efficient EVSI computation method to adjust for imperfect implementation. The implementation-adjusted EVSI is then calculated with the two methods across three examples.

\textbf{Results}: The maximum difference between the two methods remains at most 6\% in all examples. The efficient computation method is between 6 and 60 times faster than the nested simulation method in this case study and could be used in practice.

\textbf{Conclusions}: The methods developed in this paper calculate implementation-adjusted EVSI using realistic assumptions. The efficient estimation method is accurate and can estimate the implementation-adjusted EVSI in practice. By adapting standard EVSI estimation methods, we ensure that accurate adjustments for imperfect implementation can be made with the same computational cost as a standard EVSI analysis.

\textbf{Key Words}: Expected Value of Sample Information; Decision Analysis; Value of Information; Health Economic Decision Modelling; Implementation Dynamics; Research Design;

\section*{Introduction}
The Expected Value of Sample Information (EVSI) can be a tool for research prioritisation and trial design as it calculates the value of collecting additional information through a proposed study with a specific design \cite{Schlaifer:1959, RaiffaSchlaifer:1961}. When coupled with a health economic decision model \cite{Briggsetal:2006}, EVSI calculates the value of reducing the statistical uncertainty in the parameters underlying this model \emph{before} making a decision. The information collected in a study has value if it indicates that the current optimal treatment is, in fact, non-optimal. This is because the information has prevented decision-makers from implementing the incorrect treatment and thereby incurring an opportunity loss \cite{Adesetal:2004}.

EVSI can prioritise studies with the highest expected net economic benefit by computing EVSI for a range of proposed studies and subtracting the study costs \cite{ContiClaxton:2009}. This prioritisation process requires an estimate of the \emph{population-level} EVSI from which we subtract the study costs to compute the Expected Net Benefit of Sampling (ENBS) \cite{Rotheryetal:2020}. Population-level EVSI is usually estimated by multiplying the individual-level EVSI, the output of standard calculations, by the number of people who would be affected by the decision in a given year and the \emph{time horizon} of the decision \cite{Rotheryetal:2020}. This time horizon is defined as the length of time before the decision will be reassessed, i.e., due to the development of a new treatment option \cite{Philipsetal:2008}.

This estimation of the population-level EVSI assumes that any treatment recommendation made following the study is implemented instantaneously and fully \cite{AndronisBarton:2016}. In practice, this assumption is unrealistic as treatment recommendations are often slow to diffuse into clinical practice \cite{Grimmetal:2017}. Thus, standard estimates of the population-level EVSI will result in biased ENBS estimates, although the direction of this bias is dependant on the underlying decision model and the definition of the counterfactual \cite{AndronisBarton:2016, EckermannWillan:2016}. Thus, ENBS can only be estimated accurately if the population-level EVSI is adjusted for realistic expectations about the implementation of the recommended treatment, following study completion \cite{Fenwicketal:2008}.

Several frameworks can adjust for imperfect implementation, although some of these only consider the value of perfect, rather than study-specific, information \cite{Fenwicketal:2008, AndronisBarton:2016, WillanEckermann:2010, Grimmetal:2017}. Most recently, Grimm \textit{et al.} split the value of a research study into two components that focus on the research's impact on implementation and information separately \cite{Grimmetal:2017}. Within this framework, they assume that the \emph{strength} of the evidence in favour of one treatment does not influence the diffusion of the technology. However, the speed of adoption and the saturation level of the most cost-effective treatment is usually related to the strength of the evidence, which is, in turn, dependant on the future data \cite{AndronisBarton:2016}. 

Willan and Eckermann suggest that the implementation dynamics depend on the probability that a given treatment is cost effective, i.e., treatments with a higher probability of cost-effectiveness achieve a higher saturation level \cite{WillanEckermann:2010}. However, to compute the adjusted EVSI, they assume a normal distribution for the incremental net benefit distribution and the data collected in the study and obtain analytical results \cite{WillanEckermann:2010}. This restricts the use of this method to a small number of models that meet these restrictive assumptions, making it challenging to implement in practice. Thus, this manuscript extends the idea developed by Willan and Eckermann so it can be used irrespective of the underlying model structure and study design, allowing us to adjust for realistic assumptions about implementation in all EVSI calculations.

We begin by defining EVSI and demonstrating how it can be adjusted imperfect implementation  \cite{WillanEckermann:2010}. We then adapt a general purpose nested simulation algorithm to estimate the implementation-adjusted EVSI \cite{Adesetal:2004}. This algorithm can be applied irrespective of the model complexity and the data generation process. However, as is the case with non-adjusted EVSI estimation \cite{Heathetal:2020}, this method is computationally intensive, acting as a significant barrier for the proposed analyses.

Recent computation methods have been developed to efficiently compute individual-level unadjusted EVSI irrespective of the structural form of the underlying health economic decision model and study design \cite{Strongetal:2015, Menzies:2016, Jalaletal:2015, JalalAlarid-Escudero:2017, Heathetal:2017b, Heathetal:2019}. As these methods cannot directly estimate the implementation-adjusted EVSI, we present a novel adaption to one of these methods so the probability that a given intervention is cost-effective can be estimated. From this, the implementation-adjusted EVSI can then be computed. This novel method allows efficient estimation of the implementation-adjusted EVSI, based on realistic model structures and trial designs.

Following the development of these two methods, we estimate the implementation-adjusted EVSI for three proposed studies based on a previously published example. The computationally intensive nested simulation method and the adapted efficient computation method are shown to be similar. As expected, the efficient computation method is significantly faster than the nested simulation method. Importantly, by adjusting a currently available efficient computation method, we ensure that estimating the implementation-adjusted EVSI adds not further computational complexity compared to a standard EVSI analysis. Thus, the implementation-adjusted EVSI can now be easily estimated and the implementation-adjusted net value of research can be used to determine the optimal study design for future data collection, irrespective of model structure. We conclude with a discussion how these methods could be adapted to adjust for increasingly complex dynamics in the diffusion of new technologies \cite{Grimmetal:2017}.

\section*{Expected Value of Sample Information}
Health economic decision models estimate the costs and benefits of different treatment options to help decision makers select the optimal treatment from $D$ potential alternatives. These models are based on a set of model parameters $\bm\theta$ that represent real-world quantities, e.g., prevalence, quality of life weights, relative effects and treatment costs. Statistical uncertainty in the estimates of these parameters is usually characterised through a joint probability distribution $p(\bm\theta)$ in a process known as  probabilistic analysis (PA, or probabilistic sensitivity analysis, PSA). 

The costs and benefits estimated from a probabilistic health economic decision model can be combined to measure the \emph{net} benefit of a given treatment (measured in monetary or health units), denoted NB$_d(\bm\theta)$, $d = 1,\dots, D$ \cite{StinnettMullahy:1998}. Given the current evidence about the parameters, defined in $p(\bm\theta)$, the best treatment is the one that maximises the expected net benefit, $d^* = \arg\max_d \mbox{E}_{\bm\theta}\left[\mbox{NB}_d(\bm\theta)\right]$ \cite{Claxton:1999b}. The net benefit measures the average benefit across the whole population, implying that NB$_d(\bm\theta)$, $d = 1,\dots, D$, would be known if all parameter uncertainty could be resolved \cite{Briggsetal:2012}. 

EVSI calculates the value of collecting additional information about the model parameters $\bm\theta$ to improve decision making. We assume that this information is collected through a proposed research study that aims to collect data $\bm X$ \cite{Heathetal:2021}. If the data $\bm X$ were collected and realised to a specific dataset $\bm x$, then it would be combined with the current evidence to update the distribution of $\bm\theta$, $p(\bm\theta\mid\bm x)$. This updated distribution for $\bm\theta$ would change the distribution for the net benefits NB$_d(\bm\theta)$, $d = 1,\dots, D$. The (potentially new) optimal treatment would again be found by taking the expectation of the net benefit $\max_d \mbox{E}_{\bm\theta|\bm x}\left[\mbox{NB}_d(\bm\theta)\right]$. From this, the value of observing $\bm x$ is \[\max_d \mbox{E}_{\bm\theta|\bm x}\left[\mbox{NB}_d(\bm\theta)\right] - \max_d \mbox{E}_{\bm\theta}\left[\mbox{NB}_d(\bm\theta)\right].\] 

However, as the data have not been collected, we calculate the \emph{expected} value of sample information by taking an expectation over all possible studies \cite{Heathetal:2021};
\[ \mbox{EVSI} = \mbox{E}_{\bm X}\max_d \mbox{E}_{\bm\theta|\bm X}\left[\mbox{NB}_d(\bm\theta)\right] - \max_d \mbox{E}_{\bm\theta}\left[\mbox{NB}_d(\bm\theta)\right].
\]
The distribution of all potential datasets $\bm X$ is defined through the joint distribution \[p(\bm X, \bm\theta) = p(\bm X\mid\bm\theta)p(\bm\theta),\] where $p(\bm X\mid\bm\theta)$ is the sampling distribution of the data conditional on the parameters and $p(\bm\theta)$ is the current distribution of the model parameters, defined for the PA.

\subsection*{Adjusting for Imperfect Implementation}\label{EVSI-def}
We now define the implementation-adjusted EVSI (EVSI$^{IM}$) as the difference between the ``value of the current decision'' and the ``expected value of the decision made with additional information'' \cite{WillanEckermann:2010}. The value of the current decision is equal to 
\[\mathcal{C} = \sum_{d = 1}^D m_d \mbox{E}_{\bm\theta}\left[\mbox{NB}_d(\bm\theta)\right],\] where $m_d$ is the current market share of the $d$-th intervention. The value of the decision made after a specific additional dataset $\bm x$ has been collected is 
\begin{equation}
    \mathcal{F}^{\bm x} = \sum_{d = 1}^D m_d(\bm x) \mbox{E}_{\bm\theta|\bm x}\left[\mbox{NB}_d(\bm\theta)\right], \label{EVSI-IM}
\end{equation} where $m_d(\bm x)$ is the market share of the $d$-th intervention, conditional on the observed data $\bm x$. The value $\bm x$ is then the difference between $\mathcal{C}$ and $\mathcal{F}^{\bm x}$. However, the data have not been observed and so EVSI$^{IM}$ is defined by taking the expectation of $\mathcal{F}^{\bm x}$ over all possible datasets $\bm X$  \[\mbox{EVSI}^{IM} = \mbox{E}_{\bm X} \left[\mathcal{F}^{\bm X}\right] - \mathcal{C} = \mbox{E}_{\bm X} \left[\sum_{d = 1}^D m_d(\bm X) \mbox{E}_{\bm\theta|\bm X}\left[\mbox{NB}_d(\bm\theta)\right]\right] - \mathcal{C}.\] Note that $\mathcal{C}$, the value of the current decision, is unlikely to be equal to the value of the optimal treatment as we must adjust for the current issues in implementation \cite{EckermannWillan:2016}.

\subsection*{Defining the Sample Specific Market Share}
All EVSI calculation methods estimate $\mu^{\bm X} = \mbox{E}_{\bm\theta|\bm X}\left[\mbox{NB}_d(\bm\theta)\right]$, across the distribution of plausible datasets $p(\bm X)$ \cite{Kunstetal:2020}. Thus, EVSI$^{IM}$ is calculated by estimating the sample specific market share $m_d(\bm X)$. Willan and Eckermann define $m_d(\bm X)$, $d = 1,\dots, D$, as a function of the probability that a given intervention is the most cost-effective, $p_d(\bm X)$ \cite{WillanEckermann:2010};
\[m_d(\bm X) = f^m_d(p_d(\bm X)),\] with the example functional form of $f^m_d(\cdot)$ being dependant on the decision problem. For example, if clinical practice is reticent to move away from current practice ($d = 1$), then a higher probability of cost-effectiveness would be required to implement the the novel treatment ($d = 2$). Thus, to calculate $\mbox{EVSI}^{IM}$, we must estimate $p_d(\bm X)$.

\section*{Estimating the Implementation-Adjusted EVSI}
\subsection*{Nested Monte Carlo Simulation}
Unless restrictive assumptions are made, EVSI is estimated using simulation based methods \cite{Kunstetal:2020}. The first general purpose simulation method for estimating EVSI is based on a nested simulation procedure \cite{Adesetal:2004}; firstly, $S$ plausible datasets $\bm X_s$, $s = 1,\dots, S$ are simulated from $p(\bm X)$. Simulations from the marginal distribution of $\bm X$ can be obtained by simulating a parameter set $\bm\theta_s$ and then simulating a dataset from $p(\bm X\mid\bm\theta_s)$ for, $s = 1,\dots, S$\cite{Heathetal:2021}. Following this, $R$ simulations from $p(\bm\theta\mid\bm X_s)$ are required for each $s = 1,\dots,S$, denoted $\bm\theta_{r,s}$. The net benefit for each treatment must be computed for each simulated parameter set $\bm\theta_{r,s}$, $r = 1, \dots, R$, $s = 1,\dots, S$, requiring $R\times S$ evaluations of each net benefit function. The expected net benefit for treatment $d = 1,\dots, D$, conditional on the sample $\bm X_s$ is estimated by calculating the sample average net benefit \begin{equation}
    \frac{1}{R} \sum_{r = 1}^R \mbox{NB}_d\left(\bm\theta_{r,s}\right). \label{eq:mu_X}
\end{equation} The EVSI is then approximated by \[\widehat{\mbox{EVSI}} = \frac{1}{S} \sum_{s = 1}^S \max_d \left\{\frac{1}{R} \sum_{r = 1}^R \mbox{NB}_d\left(\bm\theta_{r,s}\right)\right\} - \max_d \frac{1}{S} \sum_{s = 1}^S \frac{1}{R} \sum_{r = 1}^R \mbox{NB}_d\left(\bm\theta_{r,s}\right).\]

This algorithm can be adapted to estimate EVSI$^{IM}$ by estimating the sample specific probability of cost-effectiveness for treatment $d = 1, \dots, D$ as the proportion of simulations in which treatment $d$ is optimal; \begin{equation}
    \widehat{p_d(\bm X_s)} = \frac{1}{R} \sum_{r = 1}^{R} \mathbbm{1}\left[ \mbox{NB}_d\left(\bm\theta_{r,s}\right) = \max_d \mbox{NB}_d\left(\bm\theta_{r,s}\right)\right], \label{eq:p_X}
\end{equation} where $\mathbbm{1}\left[\cdot\right]$ equals 1 if the condition is true and 0 otherwise. From this, we can estimate the market share for each treatment $\widehat{m_d(\bm X_s)} = f^m_d\left(\widehat{p_d(\bm X_s)}\right)$, and $\mbox{EVSI}^{IM}$;  \begin{equation}\widehat{\mbox{EVSI}^{IM}} = \frac{1}{S} \sum_{s = 1}^S \sum_{d = 1}^D \widehat{m_d(\bm X_s)} \left\{\frac{1}{R} \sum_{r = 1}^R \mbox{NB}_d\left(\bm\theta_{r,s}\right)\right\} - \sum_{d = 1}^D m_d \frac{1}{S} \sum_{s = 1}^S \frac{1}{R} \sum_{r = 1}^R \mbox{NB}_d\left(\bm\theta_{r,s}\right)\label{EVSI-IM-full}\end{equation} However, this algorithm is computationally intensive as it requires $S \times R$ evaluations of the $D$ net benefit functions. Thus, in practical health economic decision models where the net benefit function is non-trivial to compute, this method cannot compute $\mbox{EVSI}^{IM}$ within a feasible time frame. 

\subsection*{The Moment Matching Method}\label{ProbCE}
The Moment Matching method is an efficient nested simulation method for EVSI that reduces the number of datasets required to compute EVSI from $S$, usually at least 1000, to $Q$, which is usually between 30 and 50 \cite{Heathetal:2018}. The standard Moment Matching method approximates $\mu_d(\bm X) = \mbox{E}_{\bm\theta|\bm X}\left[\mbox{NB}_d(\bm\theta)\right]$ by reducing the variance of simulated values that derive from a function of the net benefit \cite{Heathetal:2018}. The function of $\mbox{NB}_d(\bm\theta)$ is defined by noting that the sampling distribution of $\bm X$ is typically dependent on a subset of the model parameters $\bm\phi \subset \bm\theta$. The Moment Matching method then rescales the conditional expectation of the net benefit, conditional on $\bm\phi$,  $\mbox{E}_{\bm\theta\mid\bm\phi} \left[\mbox{NB}_d(\bm\theta)\right]$. In general, $\mbox{E}_{\bm\theta\mid\bm\phi} \left[\mbox{NB}_d(\bm\theta)\right]$ can be estimated by fitting a non-parametric regression between simulated values of $\bm\phi$ and the simulated net benefit values that were calculated with the specific value of $\bm\phi$ and extracting the fitted values from this regression \cite{StrongOakley:2014}.
If the sampling distribution of the data in defined using all the model parameters then $\bm\phi = \bm\theta$, then EVSI can be approximated by rescaling $\mbox{NB}_d(\bm\theta)$ directly.

To determine the variance reduction factor, we simulate $Q$ datasets from the distribution of plausible datasets, $\bm X_q$, $q = 1, \dots, Q$ \cite{Heathetal:2021}. For each of these $Q$ datasets, we simulate $R$ values from $p(\bm\theta\mid\bm X_q)$, $\bm\theta_{r,q}$ and compute the net benefit for each treatment option, $\mbox{NB}_d(\bm\theta_{r,q})$. For each simulated dataset, we estimate the sample \emph{variance} of the net benefits, before computing the average variance across these $Q$ estimates. Finally, the simulated values for $\mbox{E}_{\bm\theta\mid\bm\phi} \left[\mbox{NB}_d(\bm\theta)\right]$ are rescaled so their variance is equal to the difference between the variance of the initial net benefit simulations and the average variance from the nested simulations \cite{Heathetal:2018}. To accurately estimate the EVSI, the $Q$ datasets should be generated so they cover the complete ``range'' of possible datasets. This can be achieved by extracting the $Q$ quantiles from the simulated values of $\bm\phi$ from the probailistic analysis. A separate dataset is then generated for each quantile \cite{Heathetal:2018}. Functions in \texttt{R} are provided in the supplementary material to estimate $\mbox{E}_{\bm\theta\mid\bm\phi} \left[\mbox{NB}_d(\bm\theta)\right]$ and specify the appropriate values of $\bm\phi$ to generate the required datasets.

\subsubsection*{Adjusting the Moment Matching Method for Imperfect Implementation}  \label{Regression}
The Moment Matching method cannot currently be used to estimate EVSI$^{IM}$ as $p_d(\bm X)$ can only be computed for the $Q$ datasets that are used in the nested simulation procedure. Thus, we now extend the Moment Matching method to compute $p_d(\bm X)$ for all plausible datasets. 

Through the standard Moment Matching method, we obtain simulations of $\mu_d(\bm X)$, across the range of plausible datasets $\bm X$. In general, $p_d(\bm X)$ will be related to the value of $\mu_d(\bm X)$, as the larger the expected net benefit, the more likely the treatment is to be cost-effective. Therefore, we now aim to estimate $p_d(\bm X)$ as a function of $\mu_d(\bm X)$. To achieve this, the nested simulations obtained during the standard Moment Matching method are used to estimate $p_d(\bm X_q)$ using equation (\ref{eq:p_X}) and $\mu_d(\bm X_q)$ from equation (\ref{eq:mu_X}), for $q = 1,\dots, Q$. From these estimates, we use non-linear regression to approximate the function  $h_d(\cdot)$; \[p_d(\bm X_q) = h_d\left(\mu_d(\bm X_q)\right) + \varepsilon_q,\] where $\varepsilon_q\sim N(0,\sigma^2)$ is the error due to estimating $p_d(\bm X_q)$ by simulation. 

We choose the functional form for $h_d(\cdot)$ by noting that (i) $p_d(\bm X)$ is a probability and thus constrained between 0 and 1, (ii) as $\mu_d(\bm X)$ increases, $p_d(\bm X)$ also increases as the treatment is becoming more valuable and (iii) $p_d(\bm X)$ increases smoothly as $\mu_d(\bm X)$ increases, i.e., if the expected net benefits are similar than the the probability of cost-effectiveness will also be similar. The generalised logistic function is a flexible function that exhibits these three features \cite{Richards:1959}, so we specify \begin{equation} h_d\left(\mu_d(\bm X)\right) = \left(A + e^{-B\mu_d(\bm X)}\right)^{-v},\label{Non-Linear-Reg}
\end{equation} where $A$, $B$ and $v$ are from the data $p_d(\bm X_q)$ and $\mu_d(\bm X_q)$ for $q=1,\dots,Q$. $A$, $B$ and $v$ can be estimated using either Bayesian or frequentist methods. However, we found that maximum likelihood methods can struggle to converge in some settings and that the all three model parameters could be estimated in a Bayesian framework using weakly informative priors to improve convergence. These priors are discussed, along with model code, in the supplementary material. 

Once estimates have been obtained for $A$, $B$ and $v$, $p_d(\bm X)$ can be estimated for each of the simulations of $\mu_d(\bm X_s)$, $s = 1,\dots, S$, denoted $p_d(\bm X_s)$. The market share of each treatment can then be computed using $m_d(\bm X_s) = f_d^m(p_d(\bm X_s)$. From this, EVSI$^{IM}$ can be estimated as;
\begin{equation}
    \widehat{\mbox{EVSI}^{IM}} = \frac{1}{S} \sum_{s = 1}^S \sum_{d = 1}^D m_d(\bm X_s) \mu_d(\bm X_s)  - \sum_{d = 1}^D m_d \frac{1}{S} \sum_{s = 1}^S \mu_d(\bm X_s)\label{EVSI-IM-MM}
\end{equation}

\subsubsection*{Estimating Implementation-Adjusted EVSI Across Sample Size}
The Moment Matching method can be extended to estimate EVSI across a range of sample sizes of the proposed study ($N_{\min}, N_{\max})$ \cite{Heathetal:2019}. This is achieved by creating a sequence of sample sizes, $N_q$, $q = 1,\dots,Q$, between $N_{\min}$ and $N_{\max}$. Each simulated dataset is then generated with a different sample size, i.e., the dataset $\bm X_q$ contains data from $N_q$ simulated individuals. The variance reduction factor for a given sample size $n$ is then found using non-linear regression. Specifically, a non-linear regression is fit between the posterior variance of the net benefit conditional on the sample $\bm X_q$ and the sample size $N_q$ \cite{Heathetal:2019}. The variance reduction factor for the sample size $n$ is then estimated by calculating the fitted value from this regression for $n$. We now extend the Moment Matching estimating method for EVIS$^{IM}$ so it can be used across sample size.

To achieve this, we proceed in a similar manner to the basic method by firstly calculating $p_d(\bm X_q)$ and $\mu_d(\bm X_q)$ for each of the nested simulations. As each of these pairs, has a different sample size, we change the regression equation so $h_d(\cdot)$ is also a function of the sample size $N_q$;
\[p_d(\bm X_q) = h_d^N(\mu_d(\bm X_q), N_q) + \varepsilon_q,\] where $\varepsilon_q \sim N(0, \sigma^2)$. In this setting, $h_d^N(\cdot)$ can still be represented by a generalised logistic function, but we adjust it to account for the fact that the larger the sample size, the \emph{faster} the probability of cost-effectiveness will increase from 0 and 1,
\[h_N(\mu_d(\bm X_q), N_q) = \left(A+e^{-B N_q^u \mu_d(\bm X_q)}\right)^{-v},\] where $u$ is an additional parameter defining the rate at which the probability of cost-effectiveness increases due to the sample size. The four parameters in this model, $B$, $u$, $A$ and $v$ can be estimated in a Bayesian or frequentist frameworks with the best performance seen using Bayesian methods with weakly informative priors.

Once these parameters have been estimated, the implementation-adjusted EVSI for a specific sample size $n$ can be estimated. To achieve this, the values for $\mu_d(\bm X)$ for the sample size $n$ are estimated using the Moment Matching method. The probability of cost-effectiveness is then estimated by computing $h_d^N(\mu_d(\bm X), n)$. From this, the market share is estimated and EVSI$^{IM}$ is calculated from equation (\ref{EVSI-IM-MM}).

\section*{Calculating EVSI Adjusted for Imperfect Implementation}
This section uses both methods to estimate EVSI$^{IM}$ for a previously developed health economic model \cite{Adesetal:2004}. We compare the two estimation methods to demonstrate their accuracy and the computational efficiency of the augmented Moment Matching method.

\subsection*{A Health Economic Decision Model for Reduced Risk of a Critical Event}
Our case study is based on a previously developed decision tree model \cite{Adesetal:2004}. This decision model has two interventions, $d = 1$ (standard care) and $d = 2$ (novel treatment). Individuals are at risk of a critical event that would lead to a reduced quality of life ($Q_C$) for the $L$ remaining years of their life and incur a yearly treatment cost ($C_C$). The novel treatment has a fixed cost ($C_T$) and reduces the probability of experiencing this critical event but with a risk of side effects. These side effects give a short term reduction in quality of life ($Q_{SE})$ and incur a one-off cost ($C_{SE}$). The model has four uncertain parameters, the baseline probability of critical event ($P_C$), the odds ratio of the critical event under the novel treatment ($OR$), the probability of side effects on treatment ($P_{SE})$ and the quality of life detriment due to the critical event. These four parameters are modelled using independent probability distributions with the distributions, and the values of the fixed parameters, given in Table \ref{tab:Parameters}. 

\begin{table}
    \centering
    \begin{tabular}{l|c|c|c}
        Description & Parameter & Mean & Distribution \\ \hline
        Probability of critical event with no treatment & $P_C$ & 0.15 & Beta(15,85)  \\
        Odds ratio of critical event with treatment & $OR$ & 0.2636 & $\log(OR) \sim N(-1.5, \frac{1}{3})$  \\
        Probability of critical event with treatment & $P_T$ & 0.0440 & $P_T = \frac{P_C OR}{1-P_c+P_C OR}$ \\
        Probability of side effects on treatment & $P_{SE}$ & 0.25 & Beta(3,9) \\
        Quality of life after critical event & $Q_C$ & 0.6405 & logit$(Q_C) \sim N(0.6, \frac{1}{6})$ \\
        Remaining years of life & $L$ & 30 & Fixed \\
        Cost of treating critical event & $C_C$ & $\$ 200,000$ & Fixed \\
        New treatment cost & $C_T$ & $\$15,000$ & Fixed \\
        Cost of treating side effects  & $C_{SE}$ & $\$ 100,000$ & Fixed \\
        Quality of life detriment due to side effects & $Q_{SE}$ & 1 & Fixed \\
        Willingness to pay for 1 quality of life unit & $\lambda$ & $\$75,000$ & Fixed 
    \end{tabular}
    \caption{The parameter specification for the decision model adapted from Ades \textit{et al.}~\cite{Adesetal:2004} and Strong \textit{et al.}~\cite{Strongetal:2015}.}
    \label{tab:Parameters}
\end{table}

The decision tree structure of the model implies that the net benefit is calculated as follows for $d = 1$ and $d = 2$, respectively:
\begin{equation*}
    \mbox{NB}_1(\bm\theta) = \lambda \left( P_C L \left(\frac{1 + Q_C}{2}\right) + (1 - P_C) L \right) - P_C C_C
\end{equation*}
\begin{align*}
    \mbox{NB}_2(\bm\theta) = \lambda \bigg( & P_T P_{SE} \left(L \frac{1 + Q_C}{2} - Q_{SE} \right)  + 
     P_T (1 - P_{SE}) L \frac{1 + Q_C}{2} + \\
    & (1 - P_T) P_{SE} (L - Q_{SE}) + (1 - P_T)(1 - P_{SE}) L \bigg) - \\
     & (C_T + P_T C_C + P_{SE} C_{SE}).
\end{align*}

\subsection*{Proposed Future Studies}
We consider three alternative proposed studies \cite{Adesetal:2004}. The first study aims to reduce uncertainty in the probability of side effects with the new treatment by offering 60 individuals the treatment and observing the number who experience side effects. The data are modelled using a binomial distribution; $X \sim Bin(60, p_{SE})$. The second study aims to reduce uncertainty in the quality of life after the critical event by recording quality of life for 100 individuals who have experienced the critical event. We model the individual level variation in the the logit of quality of life using a normal distribution with variance 2; $logit(X) \sim  N(logit(Q_E), 2)$. Finally, study 3 aims to reduce uncertainty in odds ratio of effectiveness of the new treatment compared to the standard of care. This study undertakes a randomised controlled trial with 200 patients on each arm with the data simulated from two binomial distribution, one for each treatment arm; $X_1 \sim Bin(200, P_C)$ and $X_2 \sim Bin(200, P_T)$.

\subsection*{Dynamics of Implementation}
Based on the current information, the standard care has an average net benefit of $\$2,159,300$ and the novel treatment has an average net benefit of $\$2,164,900$. Thus, the current optimal treatment is the novel treatment. However, there is substantial uncertainty about this result with only a 57\% chance that the novel treatment is the most cost-effective. To adjust for imperfect implementation, we assume that the risk of side effects for the novel treatment has made clinicians reluctant to implement it in clinical practice and it is not currently used. This means that the value of the current decision \[\mathcal{C} = \$2,159,300.\]

We then assume that clinicians will begin to adopt the novel treatment when the probability of cost-effectiveness is over 60\%. However, some clinicians will have higher levels of risk aversion and will therefore avoid the novel treatment until the evidence of cost-effectiveness is clearer. We assume that uptake of the novel treatment will be linearly related to the probability of cost-effectiveness. Furthermore, we assume that uptake will be instantaneous and static, with full treatment switching achieved if the probability of cost-effectiveness is 1. This gives a functional form of \[m_2(\bm X) = f^m_d(p_2(\bm X)) = \left\{\begin{array}{c c} 0 & p_2(\bm X)<0.6 \\ \frac{10}{4}(p_2(\bm X) - 0.6) & p_2(\bm X)\geq 0.6 \end{array}\right..\] As our example has two potential decision options, we define $m_1(\bm X) = 1 - m_2(\bm X)$ in order to estimate $\mathcal{F}^{\bm X}$ across the range of plausible datasets $\bm X$. 

\subsection*{Assessing the Performance of the Moment Matching Method}
To assess our estimation methods for $\mbox{EVSI}^{IM}$, we will calculate the implementation adjusted EVSI for the three studies. We take $S = 10,000$ to determine the expected value of the standard of care and the novel treatment under current information. To ensure a feasible computation time for the nested Monte Carlo method, we used $S = 5,000$ and $R = 10,000$. For the Moment Matching method, we took $Q = 50$ and $R = 10,000$. 

We will compare the two methods in terms of the estimates of $\mbox{EVSI}^{IM}$ across the three examples. We will also compare the estimates of the relationship between the probability that the novel treatment is cost-effective and the sample-specific incremental net benefit, i.e., the difference between the net benefit for no treatment and the novel treatment. We will compare these relationship graphically to determine whether the functional form chosen for our regression is sufficiently flexible to capture the varied relationships. Finally, we will compare the computational time required to generate the EVSI$^{IM}$ estimate.

\section*{Results}

\subsection*{Implementation-Adjusted EVSI}
Table \ref{EVSI -  Values} contains the $\mbox{EVSI}^{IM}$ estimates from the nested Monte Carlo method and the augmented Moment Matching method. The two methods are very similar with the largest discrepancy observed for Study 3, which collects additional information to estimate the odds ratio of the critical event with treatment. However, the discrepancy is only around 6\% of the EVSI$^{IM}$ estimate. Note that as both these estimates are obtained using simulation methods, some differences between the two estimates are expected. Thus, it seems likely that both methods were able to accurately estimate EVSI$^{IM}$ in this example without making restrictive assumptions about the distribution of the net benefit or the data generating mechanism.

\begin{table}
\centering
\begin{tabular}{|c|p{2.5cm}|p{2.5cm}|p{2.5cm}|p{2.5cm}|}
\hline
\multirow{2}{*}{Study} & \multicolumn{2}{|c|}{Estimate of EVSI$^{IM}$} & \multicolumn{2}{|c|}{Computational Time (s)} \\ \cline{2-5}
& Nested Monte Carlo & Moment Matching & Nested Monte Carlo & Moment Matching \\ \hline
1 - Updating $p_{SE}$ & 6086 & 6013 & 12 & 2.1 \\ \hline
2 - Updating $Q_C$ & 1924 & 1849 & 12 & 2.1 \\ \hline
3  - Updating $OR$ & 1778 & 1669 & 272 & 4.7 \\ \hline
\end{tabular}
\caption{The estimated implementation-adjusted EVSI (EVSI$^{IM}$) and the computational time required to obtain these estimates for the three studies considered for the Ades \textit{et al.} example. All estimates are obtained using both the nested Monte Carlo method and the Moment Matching method.}
\label{EVSI - Values}
\end{table}

\subsection*{Estimating the Relationship between the Probability of Cost-Effectiveness and the Expected Net Benefit}

\begin{figure}
\centering
\includegraphics[width=\textwidth]{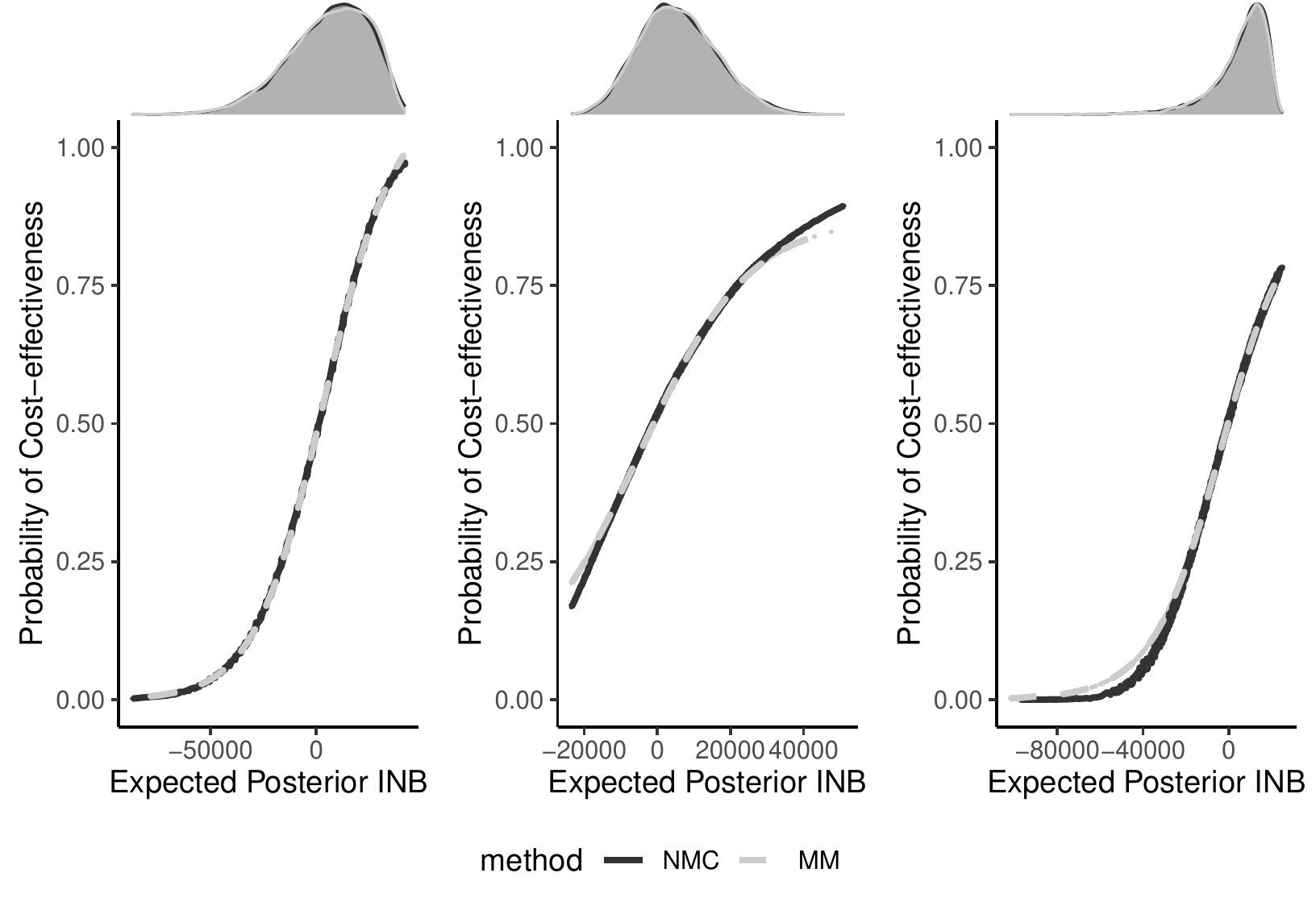}
\caption{The estimated functional relationship between the probability of cost-effectiveness and the sample specific expected incremental net benefit between the two treatment options. The black line represents the estimates generated by the nested Monte Carlo (NMC) method and the grey dashed line represents the estimated generated by the Moment Matching (MM) method. The density of the the sample specific expected incremental net benefit is represented above each plot, with a grey density plot (estimated by the MM method) plotted over a black density plot from the NMC method.}
\label{Trends}
\end{figure}

Figure \ref{Trends} plots the relationship between $p_2(\bm X)$, the probability that the novel treatment is cost-effective, and $\mu_2(\bm X) - \mu_1(\bm X)$, the expected posterior incremental net benefit. In this case, positive values of the incremental net benefit indicate that the novel treatment is optimal so you would expect $p_2(\bm X)$ to be around 0.5 when $\mu_2(\bm X) - \mu_1(\bm X)$ is equal to 0. The grey dashed line represents the relationship estimated with the Moment Matching method and the solid black line represents the relationship estimated with the nested Monte Carlo method. The relationship between $p_2(\bm X)$ and $\mu_2(\bm X) - \mu_1(\bm X)$ is similar across the two methods, for all three studies. The shape of the relationship changes across the three studies but is well captured by the generalised logistic function proposed for the regression. Discrepancies between the two curves always occur in areas of low-density for the expected incremental net benefit, shown by the density plots at the top in Figure \ref{Trends}. As EVSI$^{IM}$ is the product of $\mu_2(\bm X) - \mu_1(\bm X)$ and a function of $p_2(\bm X)$, these sections where the curve is poorly estimated have very limited impact on the overall results. These functions are displayed for a given sample size. However, the Moment Matching method can be estimated to estimate $\mbox{EVSI}^{IM}$ for different alternative sample sizes. In this case, the comparison values estimated using nested Monte Carlo would have to be recomputed and so we do not show these results. 

\subsection*{Computational Time}
Table \ref{EVSI - Values} displays the time taken (in seconds) to compute EVSI$^{IIM}$ for all three examples for the two methods. The Moment Matching method is between 6 and 60 times faster than the nested Monte Carlo method. For this example, the Moment Matching method requires 100 times fewer model runs than the nested Monte Carlo method to achieve the same accuracy. Fitting the regression model requires a fixed computational cost of around 2 seconds. Thus, in decision models where each model run has a non-negligible computational cost, the Moment Matching method will be around 100 times faster than the nested Monte Carlo method, if $S$, $R$ and $Q$ are set to the same values used in this paper.

\section*{Discussion}
It has been suggested that EVSI estimates should be adjusted to account for realistic assumptions about the implementation of new healthcare technologies \cite{WillanEckermann:2010,Fenwicketal:2008,AndronisBarton:2016, Grimmetal:2017}. Implementation of a treatment is likely to be more complete and faster if stronger evidence exists in favour of that treatment \cite{WillanEckermann:2010}. Previous, however, the only method available to compute the implementation-adjusted EVSI while taking this into account, relied on restrictive assumptions to obtain analytic formulas \cite{WillanEckermann:2010}. Thus, it was unclear how to compute EVSI$^{IM}$ in complex models that did not respect these assumptions \cite{Eckermann:2017}.

This paper addresses this gap by developing two methods to estimate the probability of the a given treatment is cost-effective across the range of plausible datasets. We develop a computationally expensive nested simulation method to estimate EVSI$^{IM}$, based on the standard nested EVSI calculation method \cite{Adesetal:2004}. We then extend the Moment Matching method for EVSI calculation to efficiently estimate EVSI$^{IM}$ by using non-linear regression to estimate the probability of cost-effectiveness from the expected posterior net benefit. These two methods provide similar estimates of EVSI$^{IM}$, although the adjusted Moment Matching method is substantially faster. We also introduce an extended Moment Matching method that compute EVSI$^{IM}$ across different sample sizes for the proposed future study.  

A limitation of this work is that to calculate EVSI$^{IM}$ using these methods, we must specify the function $f^m_d(p_d(\bm X))$ that calculates the market share based on the probability of cost-effectiveness. Our example assumed that market share increased linearly to 100\%, when the probability of cost-effectiveness for the novel treatment is over 0.6. However, $f^m_d(p_d(\bm X))$ is likely to be more complex in practice and may be challenging to determine. Grimm \textit{et al.}~used diffusion models to make realistic assumptions about the implementation changes over time \cite{Grimmetal:2017} but these would need to be re-estimated to determine how the strength of evidence impacts diffusion.

If the functional form of $f^m_d(p_d(\bm X))$ is unknown, it would be possible to undertake a sensitivity analysis to its functional form. Using these methods, this sensitivity analysis would be relatively inexpensive as the probability of cost-effectiveness would not need to be recomputed. However, it may be challenging to determine the appropriate range of functional forms that should be considered in this sensitivity analyses, especially if we wanted to allow for the possibility of implementation levels changing over time. 

Another limitation is the assumption that implementation is related to the outcome of a cost-effectiveness analysis, i.e., the probability of cost-effectiveness. Implementation could be more closely related to results based on the primary clinical outcome alone, rather than the cost-effectiveness, or based on safety concerns. These methods could be adapted to estimate the probability that a given treatment is \emph{effective}, i.e., the primary clinical outcome is largest for a specific treatment, or \emph{safe}, i.e., adverse events are lower. The market share could then be estimated based on this probability of effectiveness or safety. This analysis would jointly consider potential complementary aspects of clinical decision making, i.e., cost-effectiveness and clinical efficacy or safety, in study design. 

\section*{Acknowledgments}
The authors would like to thank Sabine Grimm for providing her helpful discussion on implementation-adjusted EVSI.

\bibliographystyle{unsrt}
\bibliography{bib}

\begin{thebibliography}{10}

\bibitem{Adesetal:2004}
A.~Ades, G.~Lu, and K.~Claxton.
\newblock {Expected Value of Sample Information Calculations in Medical
  Decision Modeling}.
\newblock {\em Medical Decision Making}, 24:207--227, 2004.

\bibitem{AndronisBarton:2016}
L.~Andronis and P.~Barton.
\newblock Adjusting estimates of the expected value of information for
  implementation: theoretical framework and practical application.
\newblock {\em Medical Decision Making}, 36(3):296--307, 2016.

\bibitem{Briggsetal:2006}
A.~Briggs, M.~Sculpher, and K.~Claxton.
\newblock {\em {Decision modelling for health economic evaluation}}.
\newblock Oxford University Press, Oxford, UK, 2006.

\bibitem{Briggsetal:2012}
A.~Briggs, M.~Weinstein, E.~Fenwick, J.~Karnon, M.~Sculpher, A.~Paltiel,
  ISPOR-SMDM Modeling Good Research Practices~Task Force, et~al.
\newblock Model parameter estimation and uncertainty: a report of the
  ispor-smdm modeling good research practices task force-6.
\newblock {\em Value in Health}, 15(6):835--842, 2012.

\bibitem{Claxton:1999b}
K.~Claxton.
\newblock {The irrelevance of inference: a decision-making approach to
  stochastic evaluation of health care technologies}.
\newblock {\em Journal of Health Economics}, 18:342--364, 1999.

\bibitem{ContiClaxton:2009}
S.~Conti and K.~Claxton.
\newblock Dimensions of design space: a decision-theoretic approach to optimal
  research design.
\newblock {\em Medical decision making}, 29(6):643--660, 2009.

\bibitem{EckermannWillan:2016}
S.~Eckermann and A.~Willan.
\newblock Expected value of sample information with imperfect implementation:
  Improving practice and reducing uncertainty with appropriate counterfactual
  consideration.
\newblock {\em Medical Decision Making}, 36(3):282--283, 2016.
\newblock PMID: 26929164.

\bibitem{Eckermann:2017}
Simon Eckermann.
\newblock {\em Health Economics from Theory to Practice}.
\newblock Springer, 2017.

\bibitem{Fenwicketal:2008}
E.~Fenwick, K.~Claxton, and M.~Sculpher.
\newblock The value of implementation and the value of information: Combined
  and uneven development.
\newblock {\em Medical Decision Making}, 28(1):21--32, 2008.

\bibitem{Grimmetal:2017}
S.~Grimm, S.~Dixon, and J.~Stevens.
\newblock Assessing the expected value of research studies in reducing
  uncertainty and improving implementation dynamics.
\newblock {\em Medical Decision Making}, 37(5):523--533, 2017.

\bibitem{Heathetal:2018}
A.~Heath and G.~Baio.
\newblock Calculating the expected value of sample information using efficient
  nested monte carlo: a tutorial.
\newblock {\em Value in Health}, 21(11):1299--1304, 2018.

\bibitem{Heathetal:2020}
A.~Heath, N.~Kunst, C.~Jackson, M.~Strong, F.~Alarid-Escudero,
  J.~Goldhaber-Fiebert, G.~Baio, N.~Menzies, and H.~Jalal.
\newblock Calculating the expected value of sample information in practice:
  Considerations from 3 case studies.
\newblock {\em Medical Decision Making}, 40(3):314--326, 2020.

\bibitem{Heathetal:2017b}
A.~Heath, I.~Manolopoulou, and G.~Baio.
\newblock {Efficient Monte Carlo Estimation of the Expected Value of Sample
  Information using Moment Matching}.
\newblock {\em Medical Decision Making}, 38(2):163--173, 2018.

\bibitem{Heathetal:2019}
A.~Heath, I.~Manolopoulou, and G.~Baio.
\newblock Estimating the expected value of sample information across different
  sample sizes using moment matching and nonlinear regression.
\newblock {\em Medical Decision Making}, 39(4):347--359, 2019.

\bibitem{Heathetal:2021}
A.~Heath, M.~Strong, D.~Glynn, N.~Kunst, N.~Welton, and J.~Goldhaber-Fiebert.
\newblock Simulating study data to support expected value of sample information
  calculations: A tutorial.
\newblock {\em Submitted to: Medical Decision Making}, -(-):--, 2021.

\bibitem{JalalAlarid-Escudero:2017}
H.~Jalal and F.~Alarid-Escudero.
\newblock {A Gaussian Approximation Approach for Value of Information
  Analysis}.
\newblock {\em Medical Decision Making}, 38(3):174--188, 2017.

\bibitem{Jalaletal:2015}
H.~Jalal, J.~Goldhaber-Fiebert, and K.~Kuntz.
\newblock {Computing expected value of partial sample information from
  probabilistic sensitivity analysis using linear regression metamodeling}.
\newblock {\em Medical Decision Making}, 35(5):584--595, 2015.

\bibitem{Kunstetal:2020}
N.~Kunst, E.~Wilson, D.~Glynn, F.~Alarid-Escudero, G.~Baio, A.~Brennan,
  M.~Fairley, J.~Goldhaber-Fiebert, C.~Jackson, H.~Jalal, et~al.
\newblock Computing the expected value of sample information efficiently:
  Practical guidance and recommendations for four model-based methods.
\newblock {\em Value in Health}, 23(6):734--742, 2020.

\bibitem{Menzies:2016}
N.~Menzies.
\newblock {An efficient estimator for the expected value of sample
  information}.
\newblock {\em Medical Decision Making}, 36(3):308--320, 2016.

\bibitem{Philipsetal:2008}
Z.~Philips, K.~Claxton, and S.~Palmer.
\newblock The half-life of truth: what are appropriate time horizons for
  research decisions?
\newblock {\em Medical Decision Making}, 28(3):287--299, 2008.

\bibitem{RaiffaSchlaifer:1961}
H.~Raiffa and H.~Schlaifer.
\newblock {\em {Applied Statistical Decision Theory}}.
\newblock Harvard University Press, Boston, MA, 1961.

\bibitem{Richards:1959}
FJ~Richards.
\newblock A flexible growth function for empirical use.
\newblock {\em Journal of experimental Botany}, 10(2):290--301, 1959.

\bibitem{Rotheryetal:2020}
C.~Rothery, M.~Strong, H.~Koffijberg, A.~Basu, S.~Ghabri, S.~Knies, J.~Murray,
  G.~Schmidler, L.~Steuten, and E.~Fenwick.
\newblock Value of information analytical methods: report 2 of the ispor value
  of information analysis emerging good practices task force.
\newblock {\em Value in Health}, 23(3):277--286, 2020.

\bibitem{Schlaifer:1959}
R.~Schlaifer.
\newblock {\em {Probability and statistics for business decisions}}.
\newblock McGraw-Hill, 1959.

\bibitem{StinnettMullahy:1998}
A.~Stinnett and J.~Mullahy.
\newblock Net health benefits a new framework for the analysis of uncertainty
  in cost-effectiveness analysis.
\newblock {\em Medical Decision Making}, 18(2):S68--S80, 1998.

\bibitem{StrongOakley:2014}
M.~Strong, J.~Oakley, and A.~Brennan.
\newblock {Estimating Multiparameter Partial Expected Value of Perfect
  Information from a Probabilistic Sensitivity Analysis Sample A Nonparametric
  Regression Approach}.
\newblock {\em Medical Decision Making}, 34(3):311--326, 2014.

\bibitem{Strongetal:2015}
M.~Strong, J.~Oakley, A.~Brennan, and P.~Breeze.
\newblock {Estimating the Expected Value of Sample Information Using the
  Probabilistic Sensitivity Analysis Sample A Fast Nonparametric
  Regression-Based Method}.
\newblock {\em Medical Decision Making}, 35(5):570--583, 2015.

\bibitem{WillanEckermann:2010}
A.~Willan and S.~Eckermann.
\newblock Optimal clinical trial design using value of information methods with
  imperfect implementation.
\newblock {\em Health economics}, 19(5):549--561, 2010.

\end{thebibliography}
\end{document}